# Waviness affects friction and abrasive wear

Yulong Li[1,2], Nikolay Garabedian[1,2], Johannes Schneider[1,2], Christian Greiner[1,2]*

[1]Institute for Applied Materials (IAM), Karlsruhe Institute of Technology (KIT), Kaiserstr. 12, 76131 Karlsruhe, Germany

[2]MicroTribology Center (µTC), Strasse am Forum 5, 76131 Karlsruhe, Germany

## Abstract

Abrasive wear can have a detrimental effect on machinery, especially in the mining and construction industries. To prolong machinery lifetime and reduce energy consumption, a thorough understanding of abrasive wear is essential: surface topography measurement and interpretation (including form, waviness, and roughness) are vitally important. However, the potentially crucial influence of surface topography intricacies on tribological behavior have been obscured since roughness and waviness are considered simple scalar quantities in most cases (e.g., roughness $R_a$ and waviness $W_t$). In this work, the complete waviness profile of the sliding track was used to shed light on the influence of surface topography on abrasive wear. Bearing steel (100Cr6, AISI 52100) pins and disks in a flat-on-flat contact were tribologically tested with $Al_2O_3$-based slurries as interfacial medium. Using slurries with two different particle sizes, 5 and 13 µm, we found that friction fluctuates only with small abrasive particles (5 µm slurry) and relatively low waviness disks. It was found that even small surface deviations (albeit minimized and controlled for) can significantly increase the friction coefficient – up to 91%. Remarkably, not only are frictional fluctuations strongly correlated with the disks' initial waviness profile, but these small fluctuations correlate with unevenly-distributed high wear. These findings enhance our understanding of the friction-wear structure and provide the basis for exploring how surfaces can be optimized for better tribological performance.

*Corresponding author: christian.greiner@kit.edu

# 1. Introduction

Undesirable abrasive wear is usually inevitable for machines, especially those operating in severe or contaminated conditions. Although researchers have produced sophisticated designs in lubrication, sealing, filtration, and other related fields, abrasive wear is still a vexing problem in practice. Aside from substantial carbon dioxide emissions due to high friction and wear, remanufacturing and replacing worn-out parts incurs high economic losses and energy consumption [1–4].

Owing to the great importance of abrasive wear, extensive studies have been performed over the past decades by both academia and industry. Researchers traditionally classify abrasive wear mechanisms into "two-body abrasion" and "three-body abrasion" based on the operating conditions of abrasive particles and their influence on wear [5,6]. Tribological performance and mechanism were found to be linked to further parameters, e.g., type of abrasive [2], abrasive size [7,8], abrasive concentration [9], and load [10,11]. Parallel wear grooves in the sliding direction and fluctuating friction have been widely reported in tribological studies as evidence of abrasive wear [12–21]. These parallel wear grooves are deemed a consequence of cutting and ploughing due to passing abrasive particles [22–25]. The source of abrasive particles could be the breakdown of the contacting surfaces, filtering failure, the malfunction of sealing, lubricant starvation, as well as numerous others. However, to our best knowledge, the cause of fluctuating friction in abrasive wear has not yet been fully understood, even if friction fluctuations are a ubiquitous manifestation of abrasive wear.

In a tribological contact, surfaces are always intricate and usually defined via form, waviness, and roughness [26] — these surface irregularities have an influence on friction and wear. The effect of roughness on friction can be positively [27], negatively [28], or non-monotonically correlated [29]. Liang et al. [30] observed a drastic fluctuation in friction coefficient with the increase of $R_a$ (average surface roughness). Due to the selection of different cut-off lengths for filtering the surface features of interest, surface roughness measurements commonly obscure the waviness portion of the surface profile [26]. Chang et al. [31] found that the waviness $W_a$ (arithmetic average of surface heights, but at higher wavelengths than simple roughness) strongly correlates with friction between Neolite rubber and quarry tiles. However, the influence of waviness on tribological contacts is not broadly studied at present and has usually been overlooked.

As shown above, in most studies, the surface profile is characterized as mean roughness and waviness parameters, which are considered simple scalar quantities (e.g., roughness $R_a$ and waviness $W_t$). These roughness and waviness parameters are not directly evaluated from the primary surface profile but generated from corresponding roughness and waviness profiles after filtering the primary profile in accordance with current standards (e.g., ISO 116610 or ASME B46.1). However, the actual surface topography is far more complicated than several scalars, which in turn may obscure the surface topography's influence on tribological behavior. Even so, these quantitative indicators describing the surface topography have to be used in the surface finishing process, as reaching a completely flat surface is almost impossible.

Given these quantification issues in surface topography and abrasive wear, we formulate a guiding question for this study: What is the systematic relationship between friction fluctuation, surface topography, and wear? In our study, instead of using quantitative indicators (e.g., $R_a$ and $W_t$), the complete waviness profile along the sliding track is used to evaluate the tribological behavior. The circular sliding track was subdivided into small segments, and the tribological data was evaluated for each of these segments and every revolution of the disk in our pin-on-disk experiments. The friction along the sliding track was then compared with the waviness profile and wear distribution. Our results strongly suggest that frictional fluctuation strongly correlates with the waviness profile and also results in uneven wear.

## 2. Experimental

### 2.1 Materials

The tribological experiments were carried out using bearing steel (100Cr6, AISI 52100) as both pin and disk material for the pin-on-disk tests. The material for the 70 mm diameter bearing steel disks were obtained from Eisen Schmitt (Karlsruhe, Germany). The disks underwent a hardening and tempering process to reach a hardness of nominally 800 HV. Surface preparation of the disks was done with a cup grinding machine (G&N MPS 2 R300, Erlangen, Germany) with corundum grinding wheels of grit EK200, resulting in reference roughness values ranging from $R_a$ = 0.08 to 0.12 µm (identical with the ISO 4287, measured by HOMMEL-ETAMIC T8000 R120-400 tactile surface profilometer, Villingen-Schwenningen, Germany). The total waviness profile height $W_t$ (based on ISO 4287) along the sliding track was specifically paid attention to (measured by FRT MicroProf optical surface profilometer MPR 1024, Bergisch Gladbach, Germany), with a reference value of $W_t ≤ 2$ µm. The pins were made by flatting and polishing 8 mm diameter spheres to a circular area of 7.33 mm in diameter. They were employed in an as-received condition with a hardness of 700 HV (purchased from KGM, Fulda, Germany). The flattened surfaces had roughness values of $R_a$ = 0.02 to 0.04 µm and a waviness $W_t$ ≤ 0.6 µm. All pins and disks were demagnetized and then cleaned with isopropanol for 15 minutes of sonication before the tribological tests.

Two kinds of water-based aluminum oxide slurries (Lapping medium BIOLAM®) with different abrasive particle sizes (5 µm and 13 µm, according to FEPA grains standard) were obtained from Joke (Bergisch Gladbach, Germany) with a concentration of 12.5 wt.%. The size distribution of the abrasive particles in the slurries was measured by laser granulometry (CILAS 1064, Orléans, France), and the result can be seen in Figure S1.

## 2.2 Tribological testing

The tribological tests were carried out with a CSEM pin-on-disk tribometer built by the Swiss Center for Electronics and Microtechnology (Peseux, Switzerland, now owned by Anton Paar). In order to segment the data from each disk revolution, a capacitive sensor and a metallic block flag were mounted on the tribometer, as shown in Figure 1a. The zero position of each circle was triggered using data from the capacitive sensor activated by the metallic block mounted on the drive shaft, as shown in Figure 1b. The tribological tests were performed at 50 mm/s with a 50 Hz sampling rate; thus, the resolution of the data acquisition is 1 mm along the sliding direction (132 mm total sliding track per revolution).

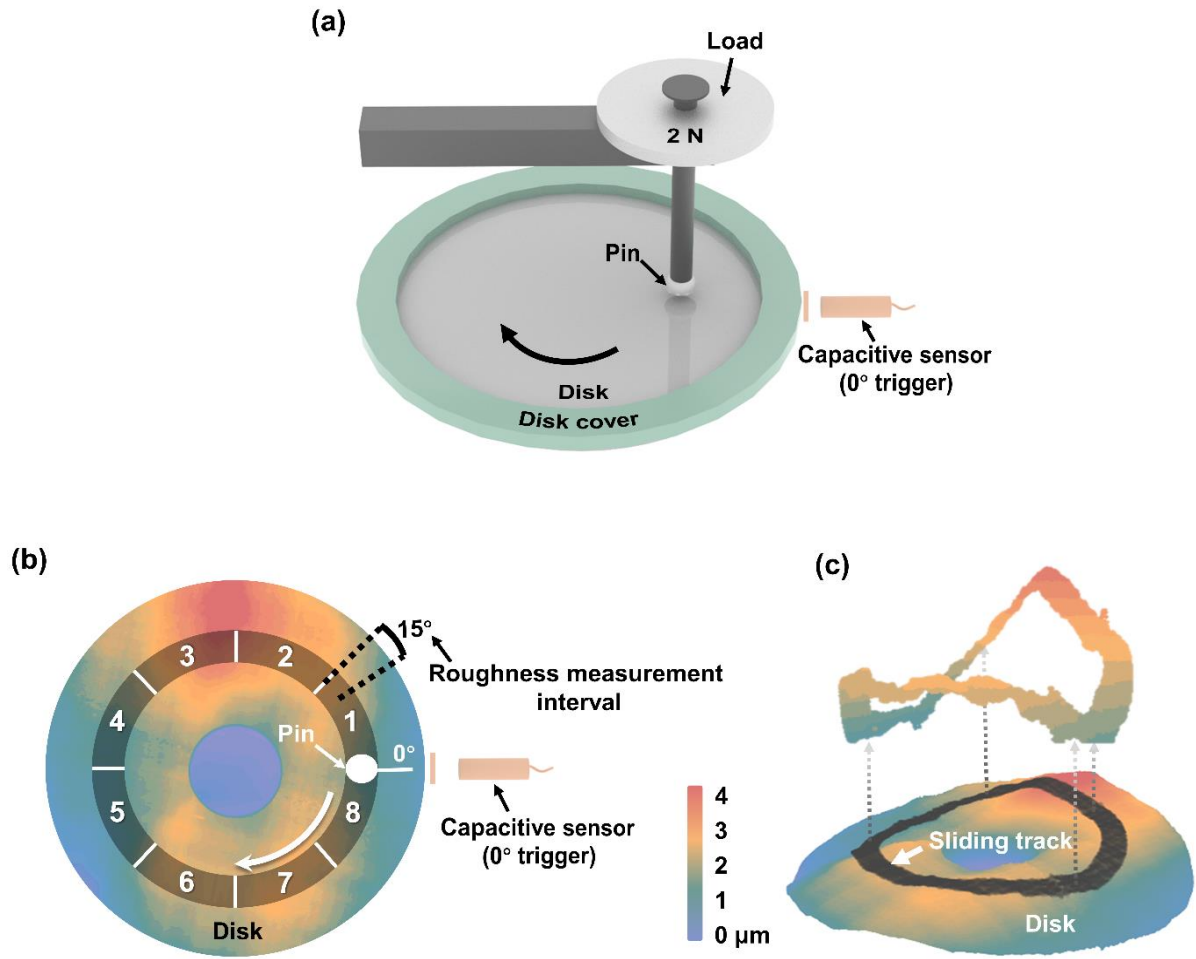

Figure 1: (a) Experimental setup of the pin-on-disk tribometer; the disk rotates clockwise; (b) Schematic illustration of the top view, separating the tribological data into 8 zones with respect to the zero-degree position defined by the capacitive sensor, and roughness measurement with every 15 degrees interval on the disks' sliding track; (c) Extracting the surface profile of the 7.33 mm width sliding track from optical surface profilometry data.

The pin was mounted in a self-aligning pin-holder for the tribological tests to ensure proper flat-on-flat contact. A dead weight of 2 N was applied on the top of the pin holder, and the friction force was measured through the deflection of an elastic loading arm. The sliding radius was set to 21 mm in order to limit the influence of a velocity gradient across the contact area [32].

Four kinds of experimental settings were tested, as shown in Table 1. Tribological tests 5_S and 5_S_50 were performed with the 5 µm slurry, while 13_S and 13_L were tested with the 13 µm slurry. 5_S, 5_S_50, and 13_S were prepared with a small waviness, where $W_t$ is less than or equal to 2 µm. 13_L has a deliberately larger waviness of $W_t \geq 7$ µm. The sliding distance was 846 m for all experiments except for one with 50 m (5_S_50). This experiment was shortened so that we could examine the surface's transient state. The experimental settings for 5_S and 13_S were repeated three times and each with fresh samples; 5_S_50 and 13_L were performed only once.

Table 1: Experiments and their settings.

|  | Slurry size | $W_t$ (Total height of the waviness profile) | Sliding distance |
|---|---|---|---|
| **5_S** | 5 µm | ≤ 2 µm | 846 m |
| **5_S_50** | 5 µm | ≤ 2 µm | 50 m |
| **13_S** | 13 µm | ≤ 2 µm | 846 m |
| **13_L** | 13 µm | ≥ 7 µm | 846 m |

**2.3 Data evaluation and characterization**

The roughness distribution was measured radially every 15 degrees of the sliding track with a stylus profilometer HOMMEL-ETAMIC T8000 R120-400 (schematic in Figure 1b). These roughness measurements were performed for all disks before and after the tribological tests. Before the tribological experiments, the measurement direction was parallel to the sliding direction. In contrast, the measurement direction was perpendicular to the sliding direction for the disks after the tests. The reason behind these choices is that we wanted to measure perpendicular to grinding/wear marks on the surfaces.

Additionally, the waviness profile and wear volume were determined via optical surface profilometry measured with a FRT MicroProf optical surface profilometer. Taking these images, first, the surface profile along the sliding track was extracted from the optical profilometry data, as shown in Figure 1c. The surface profile was then divided into 120 ring segments, with each segment covering 3° of the total circumference (width of 7.33 mm, centered at 21 mm radius). The waviness profile was evaluated from the surface profile by averaging the height of each segment and following with a profile filtering with cut-off length $\lambda_c$ = 0.08 mm (in accordance with ISO 16610). The average value of each segment is the average for the following 3 degrees. e.g., the value at x = 0° is the average of the value from 0 – 3°. The amount of wear along the sliding track is calculated by comparing the surface topography before and after the experiments.

A focused ion beam/scanning electron dual-beam microscope (FIB/SEM; Helios NanoLab Dual-beam 650 from FEI, Hillsborough, USA) was used for characterizing the worn surfaces of the disks.

## 3. Results
### 3.1 Waviness profile

Figure 2 displays the waviness profile and roughness distribution along the sliding track for four disks, which were then tested with corresponding settings presented in Table 1. There are three repeats for 5_S and 13_S, which are not discussed in the main body of paper and can be seen in the supplemental materials. For the considered specific disks with a small waviness (5_S, 5_S_50, and 13_S), the total height of the profile $W_t$ along the sliding track is 1.94 µm, 1.77 µm, and 1.56 µm, respectively. However, even with the same waviness standard ($W_t \leq 2$ µm), the waviness profiles of 5_S, 5_S_50, and 13_S differ significantly from each other. Two clear peaks can be seen for both 5_S and 5_S_50. 5_S shows a sharp peak (around 110°) considerably higher than another peak (between 230° to 300°) in 5_S, while the heights of the peaks in 5_S_50 is very close. Unlike 5_S and Disk 5_S_50, 13_S does not have significant high peaks, but several small peaks can be viewed along the sliding track. The waviness of 13_L was intentionally not mitigated during the grinding process, which yielded a total waviness profile height of $W_t = 7.24$ µm. The roughness $R_a$ along the sliding track of the disks with a small waviness (5_S, 5_S_50, and 13_S) ranges from 0.08 – 0.12 µm, with average values of 0.09 µm, 0.09 µm, and 0.11 µm, respectively. 13_L has a slightly higher average roughness, $R_a = 0.15$ µm. Similar to the waviness profiles, the roughness distribution along the sliding track varies from disk to disk.

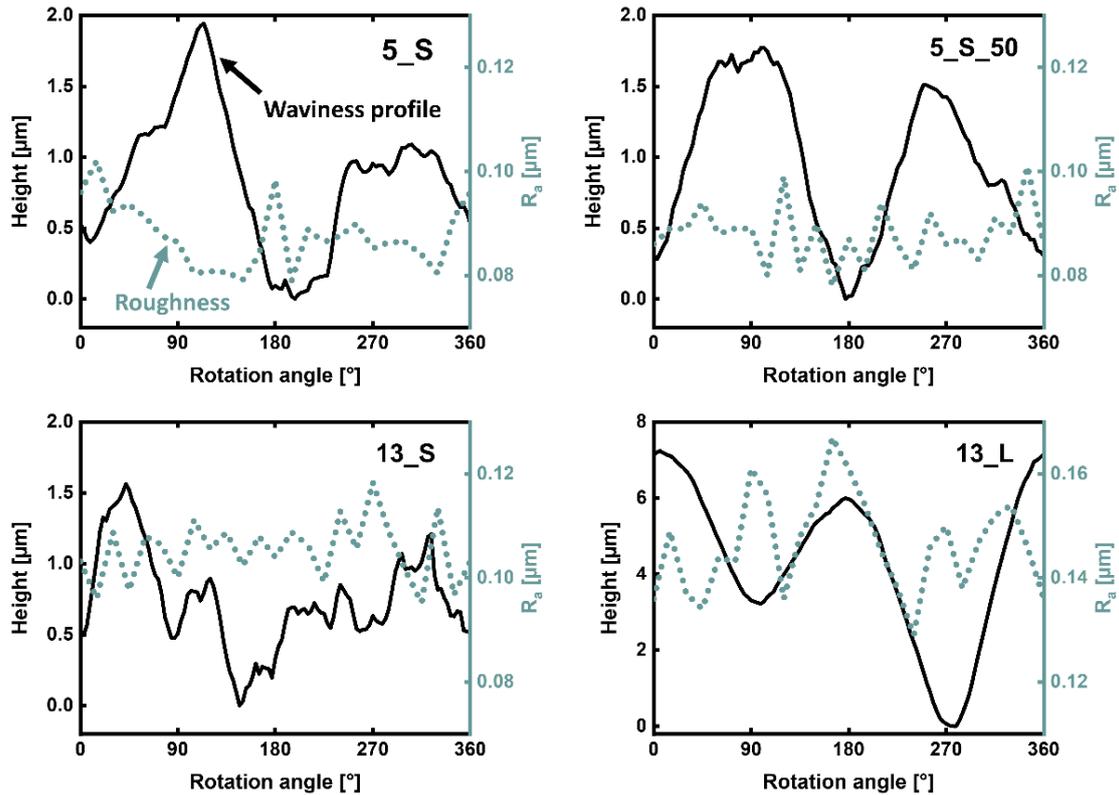

Figure 2: Waviness profile and roughness distribution along the sliding track for four individual disks: 5_S, tested with 5 μm slurry and small waviness; 5_S_50, tested with 5 μm slurry and small waviness for a short sliding distance (50 m); 13_S, tested with 13 μm slurry and small waviness; 13_L, tested with 13 μm slurry and large waviness.

### 3.2 Friction

The results for the friction coefficient as a function of the sliding distance for all tests are presented in Figure 3. No smoothing or averaging was applied to these data, and for visualization purposes only the first out of every 100 data points is displayed; the evaluations in all later parts, however, are based on the entire data obtained at a 50 Hz sampling rate. The data shown here gives an overview of the friction coefficient under abrasive wear conditions as well as the influence of the particle size. For the experiments with the 5 μm slurry, small disk waviness, and 846 m sliding distance (5_S), friction fluctuations appear from the beginning (0 – 50 m) of the test, ranging from 0.1 to 0.3. An increase in friction coefficient can be noticed after 400 m; the friction coefficient increases and then oscillates between 0.2 to 0.5. It is worth noting that although frictional fluctuations drop slightly with sliding distance, they are still considerable (around 0.2). The result of 5_S_50 (tested with 5 µm slurry and small waviness for a shorter distance of 50 m) shows a similar trend to 5_S; frictional fluctuations start from the beginning of the test, ranging from 0.1 to 0.5.

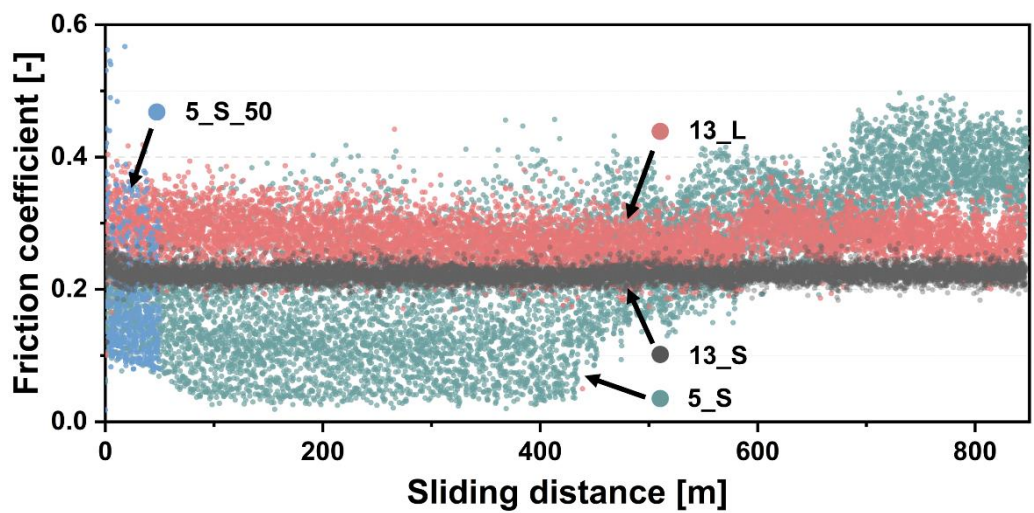

Figure 3: Friction coefficient as a function of sliding distance: 5_S, tested with 5 µm slurry and small waviness; 5_S_50, tested with 5 µm slurry and small waviness for a short distance (50 m); 13_S, tested with 13 µm slurry and small waviness; 13_L, tested with 13 µm slurry and large waviness.

The friction coefficient for the experiments with 13 µm slurry and a small disk waviness (13_S in Figure 3) exhibits a different trend than the experiments with 5 µm slurry and small waviness (5_S and 5_S_50 in Figure 3). The friction coefficient is relatively stable for 13_S, apart from a slight decrease in friction coefficient for the first 20 m, fluctuation mainly ranging from 0.2 to 0.25. The fluctuation in friction coefficient for the experiment with 13 µm slurry and a small disk waviness (13_S) is less than 1/5 compared to the tests performed with the 5 µm slurry (5_S and 5_S_50). The average friction coefficient for the experiment with 13 µm slurry and a small disk waviness (13_S) is µ = 0.22, a bit lower than that for the experiment with 13 µm slurry and increased disk waviness of around 7 µm (13_L), µ = 0.28. Moreover, 13_L has a more significant fluctuation in friction coefficient (mainly ranging from 0.2 to 0.4) than 13_S (mainly ranging from 0.2 to 0.25).

With the tribometer in Figure 1 and a sampling rate of 50 Hz, the tribological data can be divided into eight zones of 45-degree intervals (schematic in Figure 1b). Hence, it is feasible to average the friction coefficient in each zone, as shown in Figure 1b. For the result of the experiment with 5 µm slurry and a small disk waviness (5_S in Figure 4), the friction coefficient in zone 3 is relatively high (average µ = 0.33) and low friction coefficient appears in zone 1 and 8 (average µ = 0.18). Similar to 5_S, the difference in friction coefficient between the individual zones can be distinguished from 5_S_50 in Figure S2. The maximum difference in friction coefficient between each zone can be up to 0.2 for 5_S and 5_S_50, but 13_L is less prominent and here the value is only 0.04. In marked contrast to the experiment with the 5 µm slurry and a small disk waviness (5_S and 5_S_50), the friction coefficient of each zone for the experiment with 13 µm slurry and a small disk waviness (13_S in Figure 4) does show only very little difference.

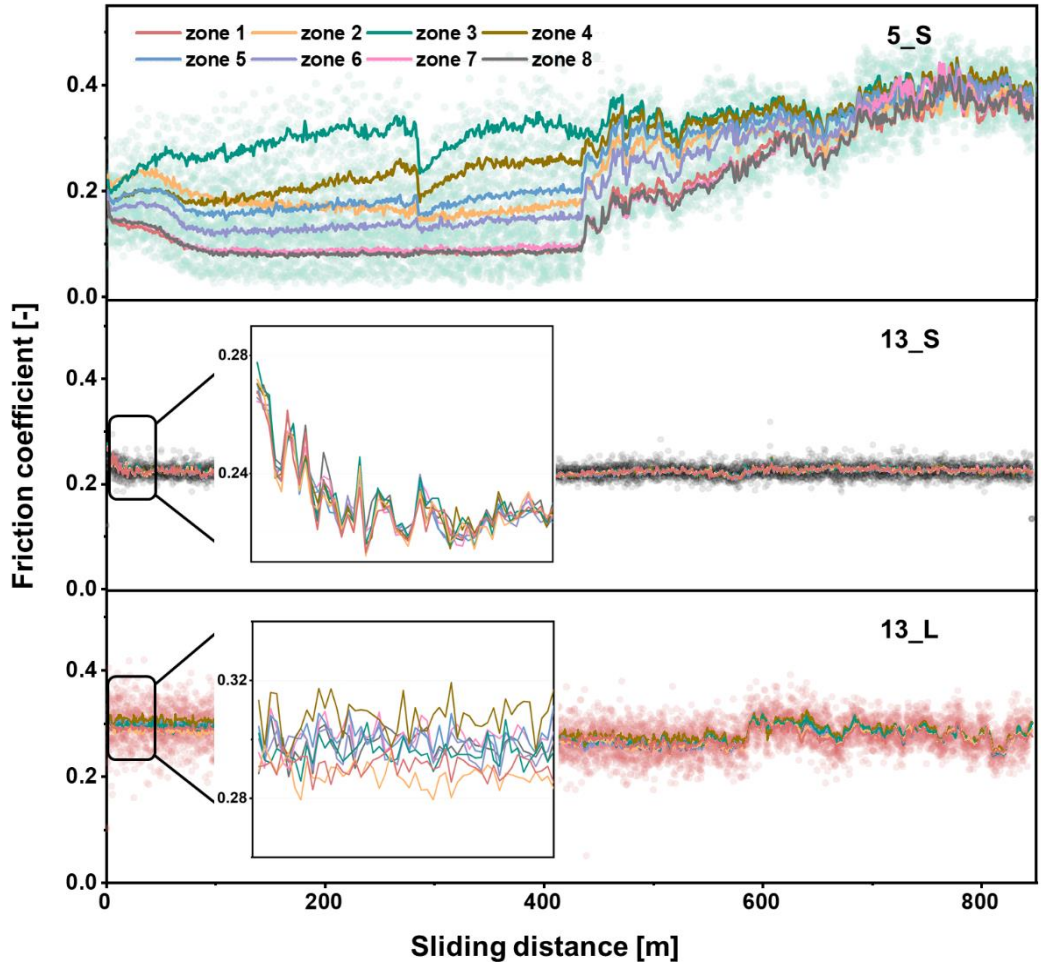

Figure 4: Friction coefficient on the disk is divided into eight zones: 5_S, tested with 5 µm slurry and small waviness; 13_S, tested with 13 µm slurry and small waviness; 13_L, tested with 13 µm slurry and large waviness.

**3.3 Wear**

In Figure 5, the wear-induced height losses for the experiments with 5 µm slurry and a small disk waviness (5_S), 13 µm slurry and a small disk waviness (13_S), and 13 µm slurry and a large disk waviness (13_L) are shown; the average wear along the sliding track are 0.39 µm, 0.31 µm, and 1.47 µm, respectively. For 13_S, the height loss on the frictional track ranges from 0.28 µm to 0.46 µm, with high wear around 0° and low wear around 180°. A significantly different behavior can be noticed for 5_S and 13_L; the wear distribution along the sliding track is uneven. For 5_S, almost no height loss is present around 0°, whereas the height loss is close to 1 µm around 110°.

Rotation angles between 60° and 160° contribute almost 2/3 of the wear. Uneven wear also occurs in 13_L: Around 150° the height loss is up to 1.93 µm, while the lowest height loss appears around 270° (about 1.15 µm).

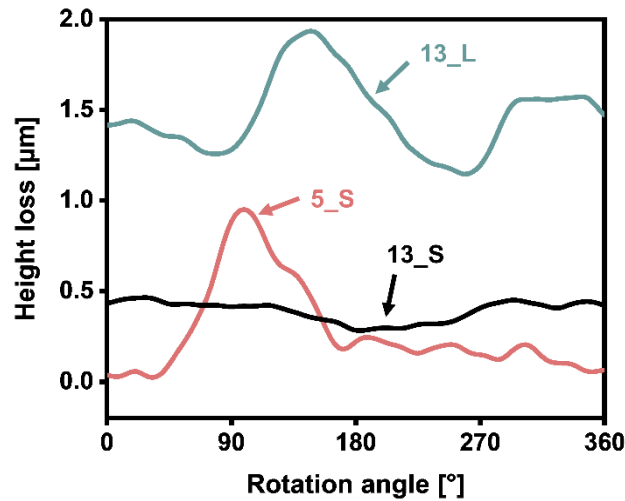

Figure 5: Wear-induced height loss along the sliding track: 5_S, tested with 5 µm slurry and small waviness; 13_S, tested with 13 µm slurry and small waviness; 13_L, tested with 13 µm slurry and large waviness.

### 3.4 Worn surfaces

Figure 6 presents SEM images of the areas with the highest and lowest friction coefficients for experiments tested with 5 µm slurry (5_S and 5_S_50) and 13 µm slurry with large waviness (13_L). All worn surfaces were imaged in the center of the wear track. For 5_S, grooves appear in both the high friction area (Figure 6a, corresponds to zone 3, around 110°) and the low friction area (Figure 6c). The grooves in the high friction area are more pronounced than those in the low friction area. The contrast in the worn surfaces of different zones also appears for 5_S_50 (Figure 6b, 6d). The test of 5_S_50 was performed for 50 m, so the vestiges of the grinding process during sample preparation still exist in the wear track (the grooves perpendicular or approximately perpendicular to the sliding direction). For the grooves generated during the tribological experiments, the high friction area (Figure 6e) of 5_S_50 has more distinct grooves than the low friction area (Figure 6b). The grooves do not have a significant difference for the high friction area and the low friction area of 13_L, in Figure 6c, f.

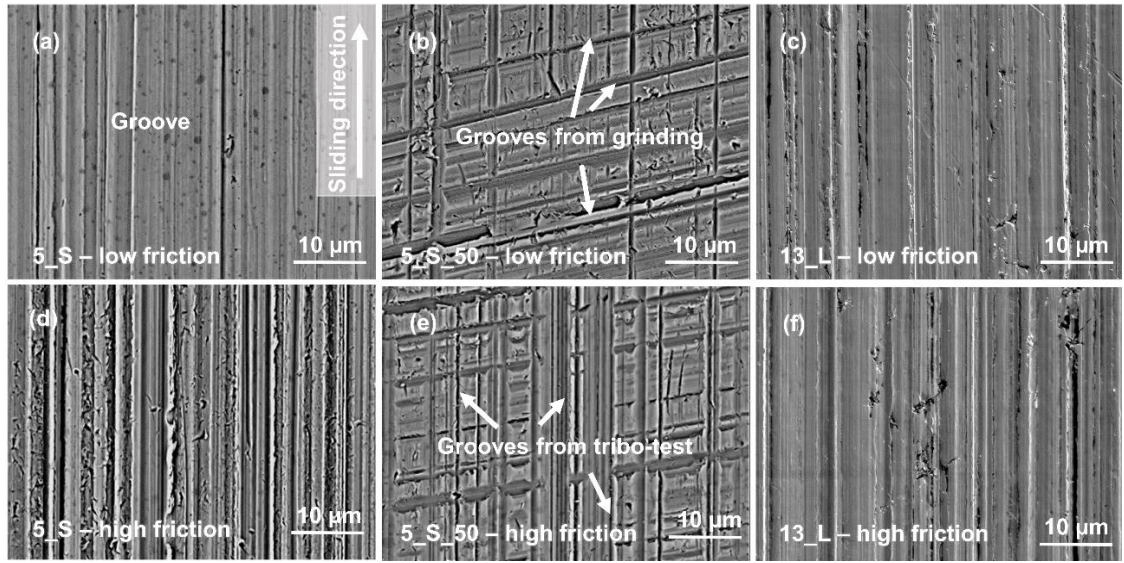

Figure 6: SEM images of worn surfaces at different areas on the sliding track: (a) and (d) tested with 5 µm slurry and small waviness (5_S), (a) is the low friction area around 0°(zone 1 and zone 8), (d) is the high friction area around 110°(zone 3); (b) and (e) tested with 5 µm slurry and small waviness for a short distance (50 m), (b) is the low friction area around 110°(zone 3), (e) is the high friction area around 300°(zone 7); (c) and (f) tested with 13 µm slurry and large waviness, (c) is the low friction area around 70°(zone 2), (f) is the high friction area around 150°(zone 4).

**3.5 Worn roughness**

The feedback between the friction coefficient and the roughness distribution along the sliding track for 5_S (tested with 5 µm slurry and a small disk waviness) and 13_L ( tested with 13 µm slurry and a large disk waviness) are highlighted in Figure 7. Here, the roughness measurements were performed on the sliding track after the tribological tests. For 5_S, the roughness $R_a$ along the sliding track ranges from 0.08 µm to 0.13 µm, with an average value of 0.10 µm. 13_L has a relatively larger roughness $R_a$ ranging from 0.19 µm to 0.23 µm, with average roughness $R_a$=0.20 µm.

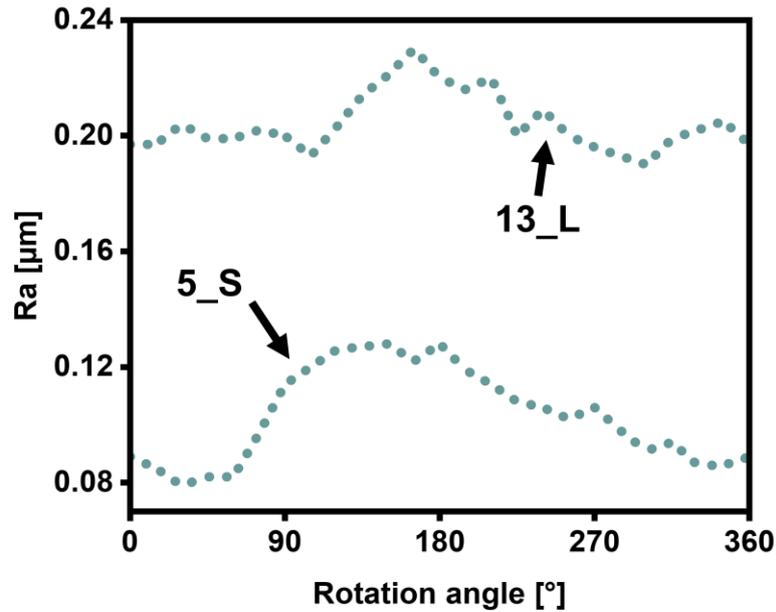

Figure 7: Roughness distribution along the sliding track after tribological test: 5_S, tested with 5 µm slurry and small waviness; 13_L, tested with 13 µm slurry and large waviness.

## 4. Discussion

In line with previous reports [16,20,21], friction coefficient fluctuations occur with abrasive wear in our study (5_S, 5_S_50, and 13_L in Figure 3). The waviness profile, roughness distribution, and wear along the sliding track were successfully obtained from optical surface profilometry data. In the discussion, the reason for the friction coefficient fluctuations is addressed first; then, the distribution of friction coefficient along the sliding track will be directly compared to the roughness distribution, waviness profile, and wear.

The tribological data of the experiments can be divided into eight zones, as shown in Figure 4. The results strongly suggested that fluctuations in friction coefficient for experiments tested with 5 µm slurry and a small waviness (5_S), 5 µm slurry and small waviness for a short distance (5_S_50), and 13 µm slurry and a large waviness (13_L) are due to inconstant friction coefficient along the sliding track. Some areas on the disk obviously lead to a higher friction, for example, zone 3 on 5_S; some areas result in a locally lower friction, e.g., zones 7 and 8 on 5_S. The inconstant friction coefficient along the sliding track on pin-on-disk setups has not been a focus of attention in most tribological studies, and few publications reported this phenomenon and even when they do, they do not point out the cause [33–35].

To elucidate what could be the reason for inconstant friction along the sliding track of 5_S, in Figure 8, the average friction coefficient along the entire wear track (in Figure 4) is compared to the roughness distribution and waviness profile (in Figure 2), respectively. The entire sliding track (360°) is divided equally into 120 segments when evaluating the average friction coefficient, and the average friction coefficient of each segment is the average for the next 3° degrees. e.g., the friction coefficient at x = 0° (as plotted) is the average friction coefficient between 0° and 3°.

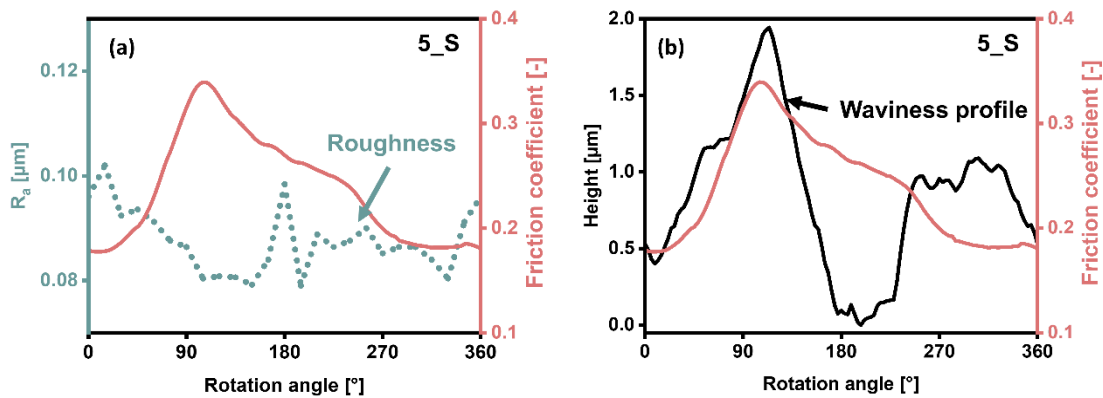

Figure 8: Comparing friction coefficient (in Figure 4) with roughness distribution and waviness profile (in Figure 2) for 5_S, tested with 5 µm slurry and small waviness: (a) friction coefficient & roughness distribution; (b) friction coefficient & waviness profile.

For 5_S (Figure 8a), there is no visible correlation between friction coefficient and roughness. By contrast, there is a partial correlation between friction coefficient and waviness profile in Figure 8b. The friction coefficient is the highest where the waviness profile has its maximum height; a "hill" of around 2 µm height can increase the friction coefficient by 91%. Similar results are obtained when testing with a 5 µm slurry and small waviness for a short distance (5_S_50 in Figure S3) and 13 µm slurry and a large waviness (13_L in Figure 9). The influence of the waviness profile on friction behavior is consistent with results published by Dai et al. [36] in molecular dynamics simulations, where they studied the molecular structure, deformation, and friction at the sliding interface between a ta-C tip and heterogeneous polymer (perfluoropolyether) surfaces; the friction force increased sharply when the probe tip overcame surface pile-ups.

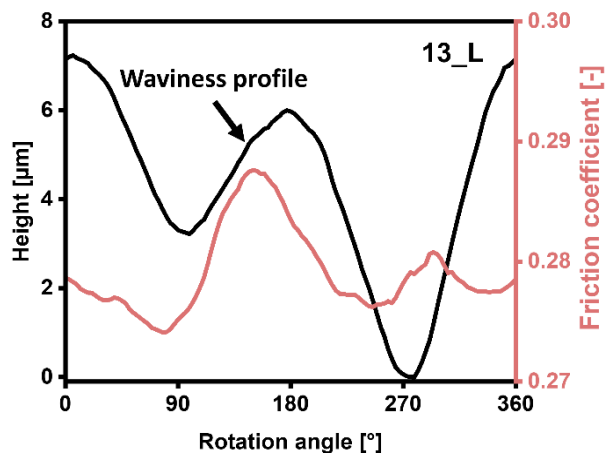

Figure 9: Comparing the waviness profile in Figure 2 and friction coefficient in Figure 4 for 13_L, tested with 13 µm slurry and large waviness.

In Dai et al.'s study, the maximum total height difference is 0.2 nm in a 30 nm sliding track (with a height-to-length ratio of 0.07). However, the ratio in our experiment is only $1.5 \times 10^{-5}$ (2 µm in 132 mm). For a steel surface, controlling the height difference to less than 2 µm on such a long sliding track (132 mm) is no easy matter. Therefore, it is astonishing to find that such a tiny height difference in the waviness profile hugely influences abrasive wear.

When making use of disks with a small waviness ($W_t ≤ 2$ µm), fluctuations in friction coefficient were only observed for experiments with the 5 µm slurry (5_S and 5_S_50 in Figure 3), while the friction coefficient in experiments with the 13 µm slurry (13_S in Figure 3) is more stable. Consistent with this observation, the results of repeated experiments for 5_S and 13_S can be found in Figure S4. The question therefore arises whether this behavior is caused by the fact that smaller slurry particles are more sensitive to the waviness profile, thereby leading to higher fluctuation in friction coefficient. With this question in mind, base bodies (e.g., 13_L) with intentionally higher waviness ($W_t = 7.24$ µm) were tested with the 13 µm slurry. By comparing the average friction coefficient along the sliding track and the waviness profile for the experiments conducted with the 13 µm slurry and increased waviness (13_L), one can infer that at locations on the disks where the waviness profile shows a positive slope (of the "hills"), the friction coefficient also increases. When $W_t$ is increased from 1.56 µm (13_S) to 7.24 µm (13_L), the base bodies' waviness profile shows an influence on the friction behavior when tested with the 13 µm slurry. This underlines that for abrasive conditions with small abrasive particles (5 µm slurry), the waviness profile has a much more pronounced influence compared to systems where the abrasive particles are larger (13 µm slurry). A critical ratio between the waviness profile $W_t$ and particle size might exist, making it even possible to predict the onset of considerable fluctuation in friction coefficient. It is therefore a possible avenue for future research in the field of abrasive wear.

In most simple terms, friction often has a strong influence on wear during tribological loading [37]; consequently, friction fluctuations might also influence wear. In Figure 5, the wear of 5_S (tested with 5 µm slurry and a small disk waviness) and 13_L (tested with 13 µm slurry and a large disk waviness) is more uneven than that of 13_S (tested with 13 µm slurry and a small disk waviness). To elucidate the relationship between fluctuations in friction coefficient and uneven wear along the sliding track, the wear-induced height loss along the sliding track (Figure 5) is directly compared to the friction coefficient for 5_S, 13_S, and 13_L (Figure 4) in Figure 10. For 5_S and 13_L, it is apparent that most of the wear occurred in the areas with the highest friction coefficient. There is a very clear and close correlation between the friction coefficient and the height loss in both cases. In contrast, for 13_S, the friction coefficient is almost constant over the entire sliding radius, and the same can – with a grain of salt – also be said about the height loss. These results indicate that fluctuations in friction, together with increased base body waviness, correlate with uneven wear along the sliding track.

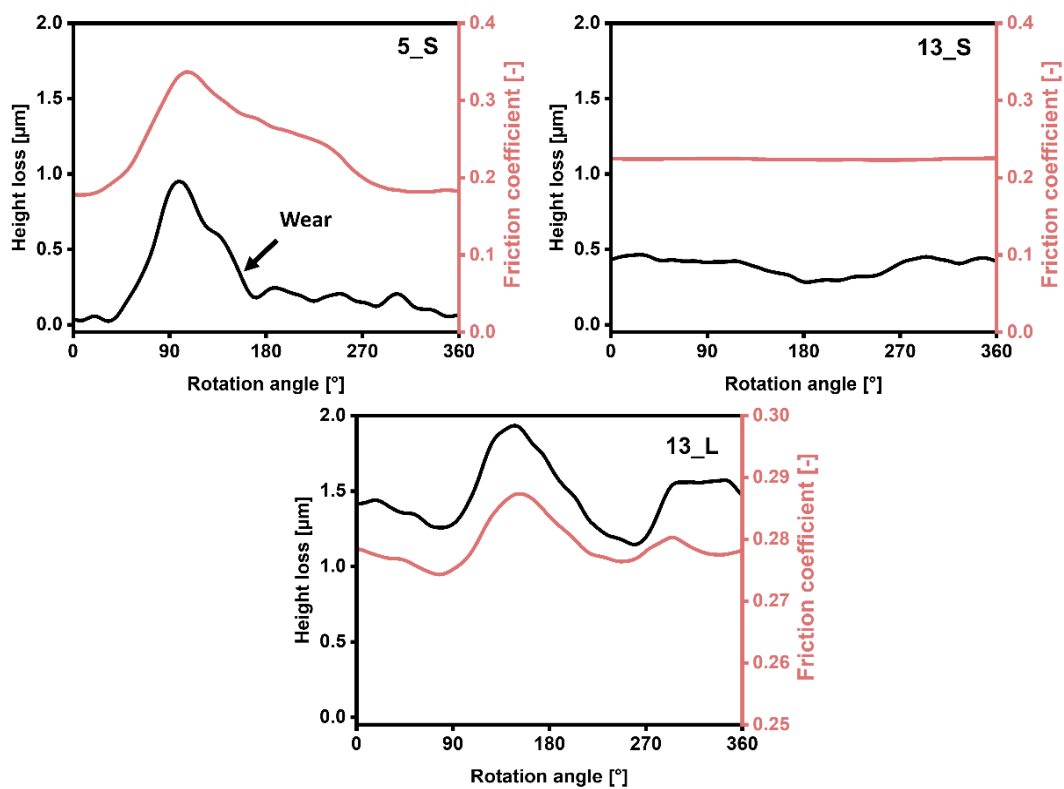

Figure 10: Comparing friction coefficient (from Figure 4) with wear-induced height loss (from Figure 5) for 5_S, tested with 5 µm slurry and small waviness; 13_S, tested with 13 µm slurry and small waviness; 13_L, tested with 13 µm slurry and large waviness.

In Figure 11, the friction coefficient for the last 50 m of the tests (790-840 m, Figure 4) and the roughness $R_a$ along the sliding track after the test (Figure 7) are compared for 5_S and 13_L. A clear correlation can be seen in this figure between in-track roughness distribution and friction coefficient. An intricate feedback mechanism seems to exist between waviness, wear, friction coefficient, and surface roughness. The SEM images of the worn surfaces presented in Figure 6 also support this line of thought. For experiments tested with a 5 µm slurry and small waviness (5_S), the portion of the disk with a higher friction coefficient (around 110°) shows more distinct grooves than the parts of the disk with a lower average friction coefficient (around 0°), which also correlates with the roughness profile presented in

Figure 11. This difference in surface morphology can be observed from early on in the tests (5_S_50, 50 m, Fig. 7b) and until the end of the experiments (5_S, 846 m, Fig. 7c). In contrast and owing to the stable friction coefficient of experiments tested with 13 µm slurry and small waviness (13_S), the worn surface in different areas shows very little difference (see Figure S5). For 13_L, the worn surfaces in high friction and low friction zones also do not show significant differences; this might be due to the fact that the friction differences in these regions are not as large as for 5_S and 5_S_50 and thereby do not cause significantly different worn surfaces. This is another indication that fluctuations in friction coefficient result in differently worn surfaces.

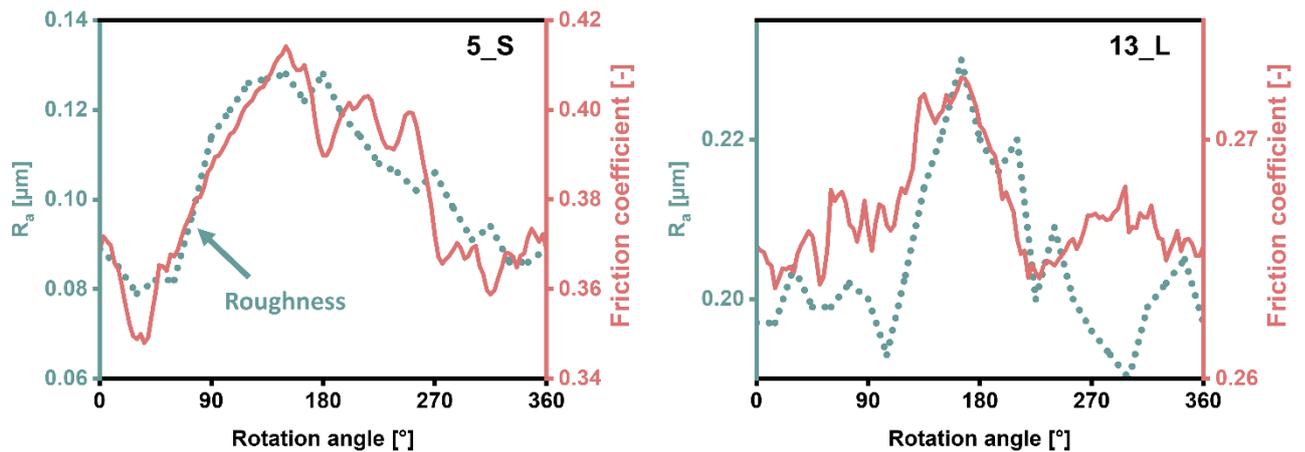

Figure 11: Comparing roughness $R_a$ along the sliding track after test (Figure 7) and friction coefficient in the later phases (790 m - 840 m in Figure 4): 5_S, tested with 5 µm slurry and small waviness; 13_L, tested with 13 µm slurry and large waviness.

Owing to the complexity of any surface profile, simple scalar quantities like $R_a$ and $W_t$ can only be first order approximations when it comes to the effect that the whole profile has in the properties of the tribological system, even if these scalar quantities (e.g., $R_a$ and $W_t$) are commonly used to define surface condition during a manufacturing process. By directly comparing the waviness profile to the tribological data, we could shed light on a previously overlooked aspect in terms of the influence of waviness on the tribological response. The data strongly suggested that apparent friction coefficient fluctuations are the result of variable friction coefficients along the sliding track, which are interrelated to the waviness profile and, eventually, lead to uneven wear. However, current surface finishing technology cannot reliably reproduce identical surface profiles, and this limits the utilization of the variable-control approach. Painting a complete picture of waviness influences [38] would require a large quantity of high-quality data which cannot be easily gathered by isolated groups and instead needs the contribution of the larger portions of the tribology community.

## 5. Conclusion

By adding a zero position trigger on a pin-on-disk tribometer, the tribological data from abrasive wear tests was separated for each revolution, and specific areas could be evaluated. The friction coefficient along the sliding track was then compared with the complete waviness profile, roughness distribution, and wear along the sliding track instead of using quantitative indicators. The data clearly demonstrates a relationship between friction fluctuations, waviness profile, and wear. Our results and their critical discussion allow us to draw the following conclusions:

- The waviness profile of the moving base body, in our case bearing steel disks, strongly influences the friction behavior. Even small deviations in the surface profile, e.g., a "hill" of around 2 µm height (along the 132 mm sliding track), can increase the friction coefficient by 91%. The local difference in friction coefficient is the reason behind the large scatter in friction data often observed in abrasive wear conditions.
- Tribological systems with smaller abrasive particles (e.g., in 5 µm slurry) are more sensitive to the waviness profile than systems with larger particles (e.g., in 13 µm slurry). As the waviness profile gets more pronounced, however, the influence of the waviness manifests itself for larger particles too.
- The local fluctuation in the friction coefficient results in uneven wear along the wear track. There is a clear, positive correlation between the local friction coefficient and the local wear.

Along with wear being affected by the waviness profile, through the friction coefficient, also the local surface roughness shows a very similar positive correlation with the waviness profile. This clearly shows the dramatic influence of the base bodies' local waviness profile on friction, wear, and surface roughness present for abrasive wear conditions; an important factor that previously seems not to have gotten the attention it deserves and that should be taken into account in the future.

## 6. Acknowledgments

We express our gratitude to the following funding agencies and collaborators: The European Research Council (ERC) under Grant No. 771237 (TriboKey). The China Scholarship Council (CSC) for awarding a scholarship to Yulong Li. The Alexander von Humboldt Foundation for awarding a postdoctoral fellowship to Nikolay Garabedian.